\documentclass[prd, twocolumn, nofootinbib]{revtex4-2}

\usepackage{graphicx}
\usepackage{dcolumn}
\usepackage{bm}
\usepackage{color}
\usepackage{amsmath}
\usepackage{hyperref}
\usepackage{aas_macros}

\hypersetup{
    colorlinks=true,
    linkcolor=red,
    citecolor=blue,
}

\newcommand{\mv}[1]{}
\newcommand{\tbd}[1]{{\color{blue} TBD}}

\begin{document}

\title{Rapid parameter estimation for pulsar-timing-array datasets with variational inference and normalizing flows}

\author{Michele Vallisneri${}^{1,2}$}
\author{Marco Crisostomi${}^{2,3}$}
\author{Aaron D. Johnson${}^{2}$}
\author{Patrick M. Meyers${}^{2}$}
\affiliation{${}^{1}$Jet Propulsion Laboratory, California Institute of Technology, 4800 Oak Grove Dr., Pasadena CA 91109}
\affiliation{${}^{2}$TAPIR, Division of Physics, Mathematics, and Astronomy, California Institute of Technology, Pasadena, CA 91125, USA}
\affiliation{${}^{3}$Dipartimento di Fisica, Universit\`a di Pisa, Largo B. Pontecorvo 3, 56127 Pisa, Italy}

\date{\today}

\begin{abstract}
In the gravitational-wave analysis of pulsar-timing-array datasets, parameter estimation is usually performed using Markov Chain Monte Carlo methods to explore posterior probability densities.
We introduce an alternative procedure that relies instead on stochastic gradient-descent Bayesian variational inference, whereby we obtain the weights of a neural-network approximation of the posterior by minimizing the Kullback--Leibler divergence of the approximation from the exact posterior.
This technique is distinct from simulation-based inference with normalizing flows, since we train the network for a single dataset, rather than the population of all possible datasets, and we require the computation of the data likelihood and its gradient.
Unlike Markov Chain methods, our technique can transparently exploit highly parallel computing platforms. This makes it extremely fast on modern graphical processing units, where it can analyze the NANOGrav 15-yr dataset in few tens of minutes, depending on the probabilistic model, as opposed to hours or days with the analysis codes used until now.
We expect that this speed will unlock new kinds of astrophysical and cosmological studies of pulsar-timing-array datasets.
Furthermore, variational inference would be viable in other contexts of gravitational-wave data analysis as long as differentiable and parallelizable likelihoods are available. 
\end{abstract}

\maketitle


Four international pulsar-timing-array collaborations recently reported evidence for a low-frequency background of gravitational waves \cite{NANOGrav:2023gor,EPTA:2023fyk,Reardon:2023gzh,Xu:2023wog}, as expected from the population of supermassive--black-hole binaries at the centers of galaxies \cite{NANOGrav:2023hfp,EPTA:2023xxk,EPTA:2023gyr,NANOGrav:2023pdq}, but also possibly from more exotic sources \cite{NANOGrav:2023hvm,EPTA:2023xxk}.
The evidence was established by testing for the gravitational-wave--specific Hellings--Downs (HD) correlations \cite{hd83} between the timing residuals of pulsar pairs across the arrays.
The tests relied on Bayesian model comparison \cite{2009MNRAS.395.1005V,taylor2021nanohertz} and on the ``optimal'' detection statistic \cite{2009PhRvD..79h4030A,2015PhRvD..91d4048C,2018PhRvD..98d4003V}: both techniques require that we obtain the posterior probability distributions $p(\theta|y)$ of the gravitational-wave and pulsar-noise parameters $\theta$ under a variety of probabilistic Gaussian-process models, which include all effects thought to significantly influence the observed residuals $y$.
The models account for pulsar geometry and kinematics, measurement noise, pulsar spin noise, dispersion in the interstellar medium, and more (see \cite{2009MNRAS.395.1005V,taylor2021nanohertz} and references therein).
Beyond the question of detection, the posteriors characterize the amplitude and spectral shape of the putative gravitational-wave signal.

\paragraph*{Approximating posteriors.} To date, posteriors have been approximated using variants of stochastic sampling (i.e., Markov Chain Monte Carlo, e.g., \cite{robert2021}), a flexible and powerful technique that nevertheless requires careful tuning; possible pitfalls include slow or poor convergence to high-probability regions of parameter space, as well as large chain autocorrelation times, which reduce the effective number of samples that carry independent information about the posterior.
It is also difficult to evaluate the appropriate length for the chains \cite{2023arXiv231102726M}.
Within the NANOGrav pulsar timing array \cite{m13,ransom+19}, the typical gravitational-wave background study relied on the \texttt{PTMCMCSampler} stochastic sampler \cite{justin_ellis_2017_1037579} to explore likelihoods computed with the \texttt{Enterprise} Python software package \cite{enterprise}, which uses \texttt{NumPy} \cite{harris2020array} and \texttt{SciPy} for the many linear-algebra operations needed for the pulsar-timing-array likelihood. \texttt{PTMCMCSampler} includes a number of specialized jump proposals to aid the exploration of parameter space.

To obtain $10^6$ chain samples, runtimes are hours for the ``CURN'' model (i.e., intrinsic pulsar red noises + Common-spectrum Uncorrelated Red Noise across all pulsars), with each likelihood taking tens of milliseconds; and days for the HD model (intrinsic pulsar red noises + a common-spectrum process with Hellings--Downs correlations), with each likelihood taking a fraction of a second \cite{2023arXiv230616223J}.
These two models have particular interest because the Bayes factor between them is the main Bayesian statistic used to establish the presence of a gravitational-wave--like signal.
Such runtimes have been barely tolerable for the NANOGrav 15-yr analysis \cite{NANOGrav:2023gor}, but they have significantly hindered the computation of detection-statistic background distributions (which require large numbers of Monte Carlo runs), and they can only grow longer for future datasets, including the joint datasets assembled by the International Pulsar Timing Array \cite{pdd+19}.
Furthermore, significantly faster parameter estimation would unlock a new class of astrophysical and cosmological analyses that are currently very challenging (e.g., joint searches for the stochastic background and multiple individual binaries; studies of background anisotropy and non-Gaussianity with many degrees of freedom; bootstrapping over pulsar subsets; and more).

While likelihoods and their stochastic exploration may yet be made faster by a combination of careful coding, more powerful processors, and smarter sampling algorithms (such as Hamiltonian Monte Carlo \cite{2011hmcm.book..113N} and Gibbs sampling \cite{2023PhRvD.108f3008L}), recent trends in high-performance computing suggest that a highly parallel scheme would yield the greatest gain.
In particular, modern high-performance graphics processing units (GPUs) have enough resources, at least on paper, to compute hundreds or thousands of pulsar-timing likelihoods simultaneously.
Unfortunately, stochastic sampling is a fundamentally serial algorithm that cannot make use of such large parallelization factors.\footnote{A possible exception may be ensemble samplers \cite{braak2006markov,goodman2010ensemble}, which however are notoriously problematic in high-dimensional parameter spaces such as ours \cite{carpenter2017}.}
Other Markov Chain Monte Carlo variants, such as parallel tempering \cite{cj1991markov}, proposal recycling \cite{452912bf-02ab-3612-90ba-770de1252be6}, and (trivially) computing multiple chains simultaneously, require smaller parallelization factors and provide more limited performance gains.

\paragraph*{Stochastic gradient-descent variational Bayes.}
In this letter we propose an altogether different method that is accurate, flexible, and ready for massive parallelization.
In \emph{variational Bayesian inference} \cite{Rezende:2015}, instead of exploring the posterior $p(\theta | y)$ we successively refine its parametrized approximation $q_\phi(\theta)$ by minimizing the Kullback--Leibler divergence of $q_\phi(\theta)$ from $p(\theta | y)$,
\begin{equation}
\label{eq:loss}
\mathcal{L}(\phi) =
\int \log \frac{q_\phi(\theta)}{p(\theta | y)} q_\phi(\theta) \, \mathrm{d} \theta.
\end{equation}
We encode $q_\phi(\theta)$ using a neural network (so that $\phi$ are the network weights), and we minimize $\mathcal{L}(\phi)$ using gradient descent---that is, by iterating $\phi \rightarrow \phi - \epsilon \nabla_{\phi} \mathcal{L}$ for an appropriate small $\epsilon$.
If the neural-network representation of $q_\phi(\theta)$ has sufficient capacity, the scheme will converge to an accurate posterior approximation as $\mathcal{L}(\phi)$ asymptotes to a constant.

The neural-network architecture of choice for this application is the \emph{normalizing flow} \cite{Tabak:2010,Tabak:2013}, in which a base distribution $q_0(x)$ (usually the unit normal $\mathcal{N}(x; 0, \mathbf{I})$, with $\mathrm{dim} \, x = \mathrm{dim} \, \theta$) is mapped into $q_\phi(\theta)$ by way of a neural network $f_\phi: x \rightarrow \theta$, so that
\begin{equation}
\label{eq:nflow}
q_\phi(\theta) = q_0\bigl(f^{-1}_\phi(\theta)\bigr)
\left| \frac{\partial f_\phi}{\partial x} \right|^{-1};
\end{equation}
the last term in the equation is the determinant of the Jacobian of the map evaluated at $x = f^{-1}_\phi(\theta)$.
To obtain a population of samples from $q_\phi$, Markov-chain style, we simply draw from $q_0$ and transform the values with $f$. 
We then obtain a Monte Carlo approximation of the loss gradient as
\begin{equation}
\label{eq:lossgrad}
\begin{aligned}
&
\nabla_{\phi} \mathcal{L}(\phi) = \frac{1}{N} \nabla_\phi \!\!\!\! \sum_{\theta^{(i)} \sim q_\phi} \!\! \left\{ \log q_\phi(\theta^{(i)}) - \log p(\theta^{(i)}|y) \right\}\\
\approx & \frac{1}{N} \!\!\!\! \sum_{x^{(i)} \sim q_0} \!\!\! \nabla_\phi \! \left\{ \log q_\phi\bigl(f_\phi(x^{(i)})\bigr)
- \log p\bigl(f_\phi(x^{(i)})|y \bigr) \right\} \\
= & \frac{1}{N} \!\!\!\! \sum_{x^{(i)} \sim q_0} \!\!\! \nabla_\phi \! \left\{ \log q_0(x^{(i)})
- \log \left| \frac{\partial f_\phi}{\partial x} \right|_{x^{(i)}} \!\!\!\!\!\!\! - \log p\bigl(f_\phi(x^{(i)})|y \bigr) \right\}.
\end{aligned}
\end{equation}
%
In the second line of Eq.\ \eqref{eq:lossgrad} we have used the \emph{reparametrization trick} \cite{2013arXiv1312.6114K} to rewrite the expectation over $q_\phi(\theta)$ as a sum over samples from the base distribution, so that the gradient $\nabla_\phi \mathcal{L}$ becomes tractable, and the sum can be parallelized trivially.
Gradient descent earns the adjective ``stochastic'' because at each iteration Eq.\ \eqref{eq:lossgrad} is evaluated over a finite \emph{batch} $\{x^{(i)}\}$ of samples.

Normalizing flows $f_\phi(x)$ are usually expressed as the composition of a number of simpler functions, and they are designed carefully to have simple Jacobians.\footnote{Analytic inverses $f^{-1}$s are also desirable to evaluate $q_\phi(\theta)$ via Eq.\ \eqref{eq:nflow}, but they are not needed for stochastic gradient descent.}
Several architectures have been studied in the literature, including real-valued non-volume preserving transformations \cite{dinh2017}, masked autoregressive flows \cite{papamakarios2017masked}, and neural spline flows \cite{durkan2019neural} which we adopt for this letter using the \texttt{flowjax} library \cite{ward2023flowjax}.

Evaluating Eq.\ \eqref{eq:lossgrad} requires an efficient algorithm to compute the likelihood and its gradient. 
To this purpose, we have reimplemented \cite{discovery} the \texttt{Enterprise} likelihood $p(y|\theta)$ using the \texttt{JAX} matrix library \cite{jax2018github}.
Functions built with \texttt{JAX} primitives can be compiled on CPUs and GPUs; they can be differentiated automatically \cite{baydin2018automatic}; and they can be parallelized to run in parallel on GPU compute cores.
These features have been specifically developed in \texttt{JAX} and similar libraries 
to enable the efficient training and evaluation of artificial neural networks.
Once we have coded $f_\phi(x)$ and $p(y|\theta)$ in \texttt{JAX}, the library will transparently provide $\nabla_\phi \log |\partial f_\phi / \partial x|$ and $\nabla_\phi \log p(f_\phi|y) = \partial p(\theta|y) / \partial \theta \times \partial f_\phi(x) / \partial \phi$, and it will automatically provide for the parallel evaluation of Eq.\ \eqref{eq:lossgrad} over a batch $\{x^{(i)}\}$.
(Beyond this parallel scheme, our \texttt{JAX} likelihood is faster than its \texttt{Enterprise} counterpart, and the availability of gradients enables methods such as Hamiltonian Monte Carlo, which we plan to explore in a separate paper.).
\begin{figure*}
    \begin{center}
    \includegraphics[width=2.25in]{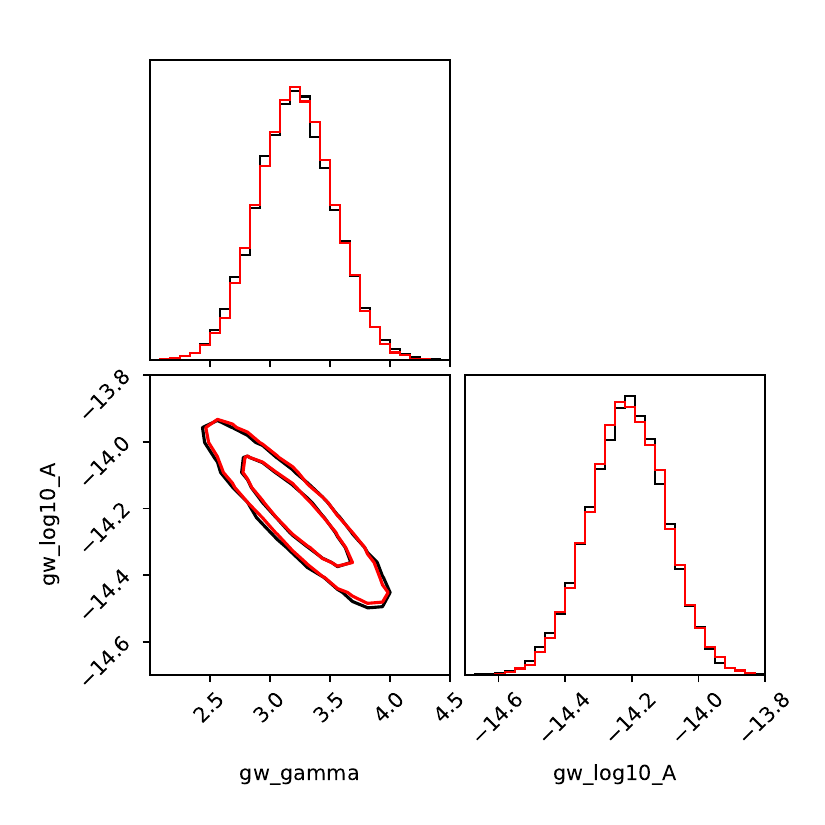}
    \includegraphics[width=2.25in]{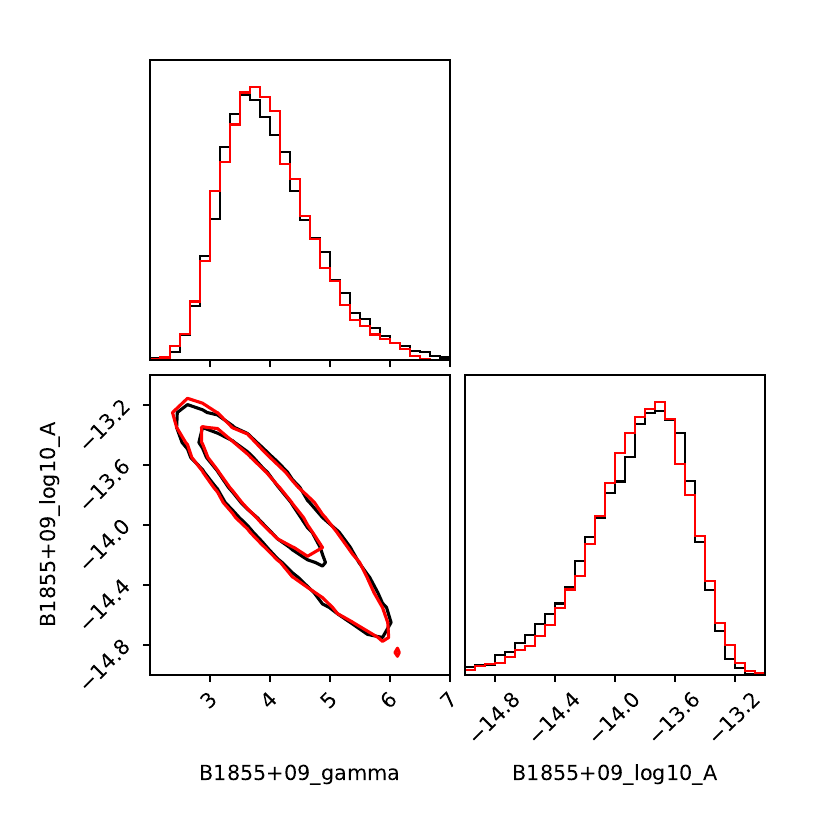}
    \includegraphics[width=2.25in]{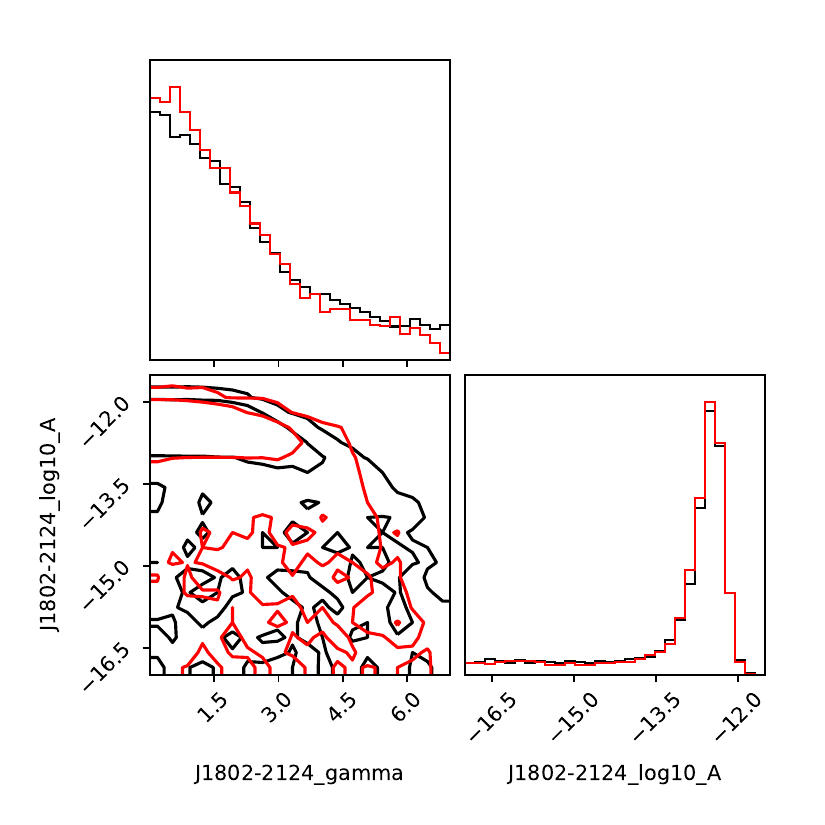}
    \end{center}
    \vspace{-12pt}
    \caption{$\mathrm{CURN}^\gamma$ $\gamma$ and $\log_{10} A$ posteriors, as evaluated using a fully trained normalizing flow (red) and No-U-Turn Hamiltonian Monte Carlo sampling with \texttt{NumPyro} (black).
    Agreement is excellent, with Hellinger distances ranging from 0.002 (for the gravitational-wave parameters) to 0.01.
    The middle panel is representative of most pulsars in the dataset; the rightmost panel demonstrates that the flow can successfully recover posterior tails for a ``difficult'' pulsar. 
    \label{fig:curn}}
\end{figure*}

Before we proceed to discuss a practical example of our scheme, we note that the normalizing-flow representation provides also an expedient importance-sampling estimator of the marginal likelihood (a.k.a.\ Bayesian evidence) as \cite{rezende2014stochastic, Srinivasan:2024uax}
\begin{equation}
\int p(y | \theta) p(\theta) \, \mathrm{d}\theta
\approx \frac{1}{N} \!\! \sum_{\theta^{(i)} \sim q_\phi} \frac{p(y | \theta^{(i)}) p(\theta^{(i)})}{q_\phi(\theta^{(i)})},
\end{equation}
where the sum can be taken over $x^{(i)} \sim q_0$, and the denominator evaluated via Eq.\ \eqref{eq:nflow}.


\paragraph*{Examples.} 
We perform variational inference on a simulated dataset modeled on NANOGrav's ``15-yr'' data release: 67 pulsars timed over 16 years, with $\sim$ 675,000 timing residuals \cite{2023ApJ...951L...9A,15yrdataset}. 
We estimate posteriors for the $\mathrm{CURN}^\gamma$ and $\mathrm{HD}^\gamma$ probabilistic models of Ref.~\cite{NANOGrav:2023gor}, which share a common power-law Gaussian process to represent gravitational waves, but differ in the presence of Hellings--Downs correlations.
Each model is fully parameterized by the red-noise log amplitudes $\log_{10} A_k$ and spectral slope $\gamma_k$ (where $k$ ranges over the array pulsars), and 
by the gravitational-wave $\log_{10} A_\mathrm{gw}$ and spectral slope $\gamma_\mathrm{gw}$, for a total of 136 parameters.
Our simulation is obtained by using $\mathrm{HD}^\gamma$ as a generative probabilistic model, setting parameters in the vicinity of their maximum-likelihood values in the NANOGrav data release.

Our $f_\phi$ is a \emph{neural spline flow} \cite{durkan2019neural} with 16 flow steps and 8 bins, for a total of $\sim$ 365,000 network parameters.
We train the $q^\mathrm{CURN}_\phi$ approximant over batches of 3,584 normal draws.
Each batch evaluation takes 0.4 s on a 40-GB A100 NVIDIA GPU, using double-precision floating-point math; after a few hundred iterations the loss begins to flatten out and posterior contours stabilize.
Full convergence of the loss is reached at $\sim$ 2,200 iterations (about 15 minutes).
Since each training batch includes newly drawn samples from the base distribution, overfitting is not a concern here.
Convergence can be accelerated (although we did not do so here) by \emph{annealing} the posterior so that it begins as a softer target for the approximant. This is achieved in practice by multiplying the $\log p$ term in Eq.\ \eqref{eq:lossgrad} by a factor that grows from 0 to 1 over the first few hundred iterations.

The resulting $\mathrm{CURN}^\gamma$ posteriors for the gravitational-wave and red-noise parameters are shown in Fig.\ \ref{fig:curn}, compared to 16,384 posterior sample obtained with the ``No-U-Turn'' Hamiltonian Monte Carlo sampler \cite{hoffman2014no} in its \texttt{NumPyro} implementation \cite{phan2019composable,bingham2019pyro}, again using our JAX likelihood.
The one-dimensional Hellinger distance \cite{Hellinger1909} between the distributions is $\sim 0.01$ for most parameters and $0.002$ for the gravitational-wave parameters; the latter number is comparable to the distance between subsamples from Hamiltonian Monte Carlo.

On the A100, the $q^\mathrm{HD}_\phi$ approximant can be trained over smaller batches of 256 normal draws, because the $\mathrm{HD}^\gamma$ likelihood is more memory intensive.
Each evaluation takes 1.2 s; training converges fully after $\sim$ 1,200 iterations (about 25 minutes, although the JAX compilation of the HD loss and loss gradient takes several minutes, as opposed to seconds for CURN.)
Parallel evaluation is limited by the number of GPU compute cores and by the available GPU memory, so training will be faster with even more capable cards.
It is likely that the network architecture used here could be optimized further for both speed and accuracy. Nevertheless, for both models the normalizing-flow distributions appear to be very accurate approximations of the true posteriors.

Several avenues are possible to verify the accuracy of results without reference to independent stochastic runs. In addition to confirming that the training loss has converged and that the final posteriors are independent of the pseudorandom-number seeds used in training, one may probe whether network capacity is sufficient by comparing posteriors obtained by changing the number or size of the network layers; and one may refine the converged $q_\phi$ using \emph{posterior reweighting} \cite{2023PhRvD.107h4045H,2023PhRvL.130q1403D}, in which accurate approximations would be seen to yield a small weight variance and a large effective number of samples \cite{elvira2022}.
We plan to perform a detailed quantitative analysis of normalizing-flow capacity, training convergence, and posterior accuracy in a follow-up paper.

\paragraph*{Discussion.}
Stochastic gradient-descent variational Bayes with normalizing flows seems optimally adapted for pulsar-timing-array data analysis, in which we deal with a single dataset, and we can now efficiently evaluate the likelihood and its gradient in parallel on GPUs.
The scheme provides a generative representation of the posterior (i.e., it can draw posterior samples very rapidly), but it can also evaluate the posterior at any parameter location (with the correct normalization if the likelihood itself is normalized), and it can approximate the marginal likelihood.

We note that normalizing flows have been used to great effect in gravitational-wave data analysis \cite{Green:2020hst, Green:2020dnx, Dax:2021tsq, Dax:2021myb, 2023PhRvL.130q1403D, Wildberger:2022agw, Bhardwaj:2023xph, Crisostomi:2023tle, Leyde:2023iof}, including pulsar timing arrays \cite{Shih:2023jme}, to implement ambitious \emph{simulation-based-inference} schemes whereby one approximates the posterior density $p(\theta|y)$ \emph{as a function of $y$} by training the network using a large number of simulated datasets $\{\theta^{(i)},y^{(i)}\}$ and a loss function different from ours.
These schemes offer \emph{amortized} inference: the network is trained \emph{a priori}, with large computational cost, by considering all possible presentations of noise and signal; parameter inference with the actual observed data is then almost instantaneous.
By contrast, in our scheme the network is trained anew for each dataset; however, given that we have to learn only one posterior, the $f_\phi$ network can be much simpler, and can be optimized more cheaply.

As we have shown, our algorithm can already handle realistic datasets under the standard probabilistic models that have provided evidence for low-frequency gravitational waves \cite{NANOGrav:2023gor}.
The scheme offers the so far unrealized prospect of \emph{real-time Bayesian inference}, which may be especially beneficial as we evaluate data-reduction or noise-modeling problems with individual pulsars, or as we explore the design space of probabilistic models.
In fact, we expect that the networks that we have demonstrated here would easily extend to the hierarchical Bayesian models advocated by van Haasteren \cite{vh24}, which introduce population priors for the intrinsic red-noise hyperparameters. These models could also be explored by reweighting normalizing-flow samples by the ratio of hierarchical and original red-noise priors.

Another intriguing possibility would be representing single-pulsar posteriors with trained normalizing flows in a manner similar to Ref.\ \cite{2022PhRvD.105h4049T}, and then exploring joint array posteriors by imposing common priors on the single-pulsar parameters, as suggested by Lamb and colleagues \cite{2023PhRvD.108j3019L}. Unlike the kernel density estimators of Ref.\ \cite{2022PhRvD.105h4049T}, the normalizing-flow posteriors would account fully for correlations.
The modular machine-learning architecture of this scheme will make it possible to experiment with promising innovations in normalizing flows and more generally in neural networks, and to take advantage of machine-learning paradigms such as \emph{transfer learning} \cite{bozinovski2020reminder} (for instance, training a $\mathrm{HD}^\gamma$ flow more quickly by starting from the converged parameters of a $\mathrm{CURN}^\gamma$ flow).
Last, there may be opportunities to apply the scheme to other use cases in gravitational-wave data analysis where we seek rapid parameter estimation under a range of models and we can deploy differentiable and parallelizable likelihoods \cite{ripple}. 

\vspace{12pt}
\paragraph*{Acknowledgements.}
We are grateful to Katerina Chatziioannou, Maura McLaughlin, Rutger van Haasteren, and Stephen Taylor for useful discussions and valuable comments on our draft.
We acknowledge support from National Science Foundation (NSF) Physics Frontiers Center award numbers 1430284 and 2020265 and from the Jet Propulsion Laboratory President's and Director's Research and Development fund.
M.C. is funded by the European Union under the Horizon Europe's Marie Sklodowska-Curie project~101065440.
This research was performed at the Jet Propulsion Laboratory, California Institute of Technology under a contract with the National Aeronautics and Space Administration.
Copyright 2024. All rights reserved.


\bibliography{vbayes}

\end{document}